\documentclass[10pt,conference,compsocconf]{IEEEtran}

\usepackage[ruled,vlined]{algorithm2e}
\usepackage{amsfonts}
\usepackage[cmex10]{amsmath}
\usepackage{amssymb}
\usepackage{amsthm}
\usepackage{array}
\usepackage{booktabs}
\usepackage{cite}
\usepackage{color}
\usepackage{dsfont}
\usepackage{float}
\usepackage[T1]{fontenc}
\usepackage{balance}
\usepackage{graphicx}
\usepackage[utf8]{inputenc}
\usepackage{mathtools}
\usepackage{multirow}
\usepackage[all]{nowidow}
\usepackage{tikz-cd}
\usepackage[hyphens]{url}

\usepackage{paralist} 
\let\labelindent\relax 
\usepackage{enumitem}

\begin{document}

\title{Deadline-Aware Multipath Communication:\\ An Optimization Problem}

\author{
	\IEEEauthorblockN{Laurent Chuat\IEEEauthorrefmark{1}, Adrian Perrig\IEEEauthorrefmark{1}, Yih-Chun Hu\IEEEauthorrefmark{2}}
	\IEEEauthorblockA{\IEEEauthorrefmark{1}Department of Computer Science, ETH Zurich, Switzerland}
	\IEEEauthorblockA{\IEEEauthorrefmark{2}Department of Electrical and Computer Engineering, University of Illinois at Urbana-Champaign, USA}
}

\maketitle

\begin{abstract}
Multipath communication not only allows improved throughput but can also be used
to leverage different path characteristics to best fulfill each application's
objective. In particular, certain delay-sensitive applications, such as
real-time voice and video communications, can usually withstand packet loss and
aim to maximize throughput while keeping latency at a reasonable level. In such
a context, one hard problem is to determine along which path the data should be
transmitted or retransmitted. In this paper, we formulate this problem as a
linear optimization, show bounds on the performance that can be obtained in
a multipath paradigm, and show that path diversity is a strong asset for
improving network performance. We also discuss how these theoretical limits can
be approached in practice and present simulation results.
\end{abstract}

\section{Introduction}
\label{sec:intro}
Looking back in history, many computer systems were initially designed to use
only a single resource of each type~(e.g., processor, memory, display) at once.
Over time, to increase the performance of these systems, we have seen the
development of better components (in terms of speed, capacity, or size) and more
efficient algorithms; but these options are limited by the laws of physics, the
ingenuity of researchers, and the nature of the problem. \emph{Parallelism}
emerged as the other options faced barriers (as illustrated by the development
of multicore processors, for example). Oddly, the idea of applying parallelism
to network paths (i.e., using multipath protocols) has only started to get
traction recently, with Multipath TCP~\cite{Raiciu2012} in particular.

There is an ongoing effort to develop new network protocols and
improve existing ones, but even with an optimal communication strategy,
performance depends on the underlying medium of data transfer. Among the
different physical means of carrying data that we know of (e.g., optical fiber,
electromagnetic radiation, or a pair of conductors), there is no panacea. Fiber
unquestionably allows greater throughput than copper wires, for example,
but microwaves offer an important advantage for mobility
and can significantly cut latency. In fact, the speed of microwaves on the
surface of the earth is close to the speed of light in vacuum, whereas fiber
only achieves roughly $2/3$ of that speed~\cite{Singla} (even when assuming a
straight line between source and destination). However, these benefits
come at a cost, namely higher loss rates (which depend on distance
and other environmental conditions) and lower bandwidth.

In a near future, we might witness the appearance of even more heteroclite
networks with projects such as Facebook's Aquila~\cite{Aquila} (based on
solar-powered drones), Google's Project Loon~\cite{Loon} (based on high-altitude
balloons), or SpaceX's project to provide Internet with low-orbit
satellites~\cite{SpaceX}. Furthermore, future Internet architectures could
explicitly provide multiple paths to end hosts~\cite{ZhHsHaChPeAn2011, Nebula,
segmentrouting, YanClaBer2007, GoGaShSt2009} and these paths might also
exhibit very diverse properties. As sending multiple packets over
a network in which multiple paths are available is a parallelizable task by
nature, we claim that it is possible to take advantage of this situation to
accomplish a  broad set of application-level objectives, such as latency-related
objectives. Unfortunately, as of today, few protocols make use of multiple
network paths simultaneously.

Because most applications are sensitive to latency to some extent, we
distinguish between two main classes of applications. The first class concerns
applications that need a reliable transport protocol---typically TCP or a
variant thereof~(possibly with multipath functionalities and/or optimized for
latency, see Section~\ref{sec:related}). This class encompasses file transfers,
web browsing, and more. For these applications latency might be an important
concern, but reliability is the critical requirement. The second class relates
to real-time applications, which typically do not use a reliable protocol.
The reasons for not using a fully reliable service are the following.
By definition, reliable protocols never discard any packet before the sender
receives an acknowledgment---even if the packet in question is obsolete from the
application's perspective. Moreover, ordered byte-stream protocols suffer from
the head-of-line blocking problem, and cannot give any guarantee regarding the
time at which a packet will be delivered. As a consequence, it is hard to
specify strict latency-related objectives when a reliable protocol such as TCP
is considered; hence real-time applications typically use UDP instead. The
problem when using UDP is that transport-layer duties (e.g., retransmissions,
congestion control) are delegated to the application layer, thus every
application must re-implement the same mechanisms. Furthermore, as of today,
multipath functionalities are not natively available to the developers of such
applications.

In this paper, we focus on real-time applications and consider one particular
communication scenario in which the objective is to deliver as much data
as possible before a \emph{deadline} across multiple end-to-end paths.
After the deadline, the data can be discarded (i.e., the
communication is not fully reliable, but gives latency guarantees). There is a
plethora of applications that would benefit from a deadline-aware protocol:
voice communication, videoconferencing, live video streaming, online gaming,
high-frequency trading, and more. The lifetime of a
packet for these various applications could range from a few milliseconds to
several seconds.\footnote{Latencies of 20--30 \emph{ms} are considered as relatively
high, although acceptable, for musical applications; and humans can tap a steady
beat with variations as low as 4 \emph{ms}~\cite{Lago2004}. On the other hand,
live YouTube streams can be broadcast with latencies on the order of
seconds~\cite{YouTubeStreaming}.} Therefore, it is crucial that practical
techniques as well as theoretical foundations be developed for partially
reliable multipath communications.

Using multiple paths simultaneously implies that the sender might have to make
hard decisions regarding packet-to-path assignments when the available paths
have different properties. It may not be obvious that path diversity can help
improve network performance. Is it preferable to have identical paths (in which
case the packet assignment problem becomes irrelevant) or diverse ones? Also, if
diverse paths are available, is the optimal strategy to always use only one of
these paths (the most appropriate one from the application perspective)?
Intuitively, diversity allows each path to specialize in a different task.
High-bandwidth paths can carry the initial data transmission, and low-latency
low-loss paths present advantages for retransmissions and control data (e.g.,
acknowledgments). We provide a model that allows determining the
potential benefits of any given set of paths and we show in our evaluation that
having complementary paths is beneficial in a deadline-based communication
context to.

\section{Problem Description}
\label{sec:problem}
One typical situation in which two paths are available is when a smartphone is
connected to both a WiFi access point and a cellular network. This can lead to
very different outcomes depending on which path is selected. Bandwidth depends
on which generation of the technology in question is used (e.g., 3G, 4G,
802.11a, b, g). Losses depend on congestion, environmental conditions, and more.
Finally, delay is in part influenced by the signal quality as retransmissions
can be performed at the link layer.

\begin{figure}[h!]
\centering
\includegraphics[width=\columnwidth]{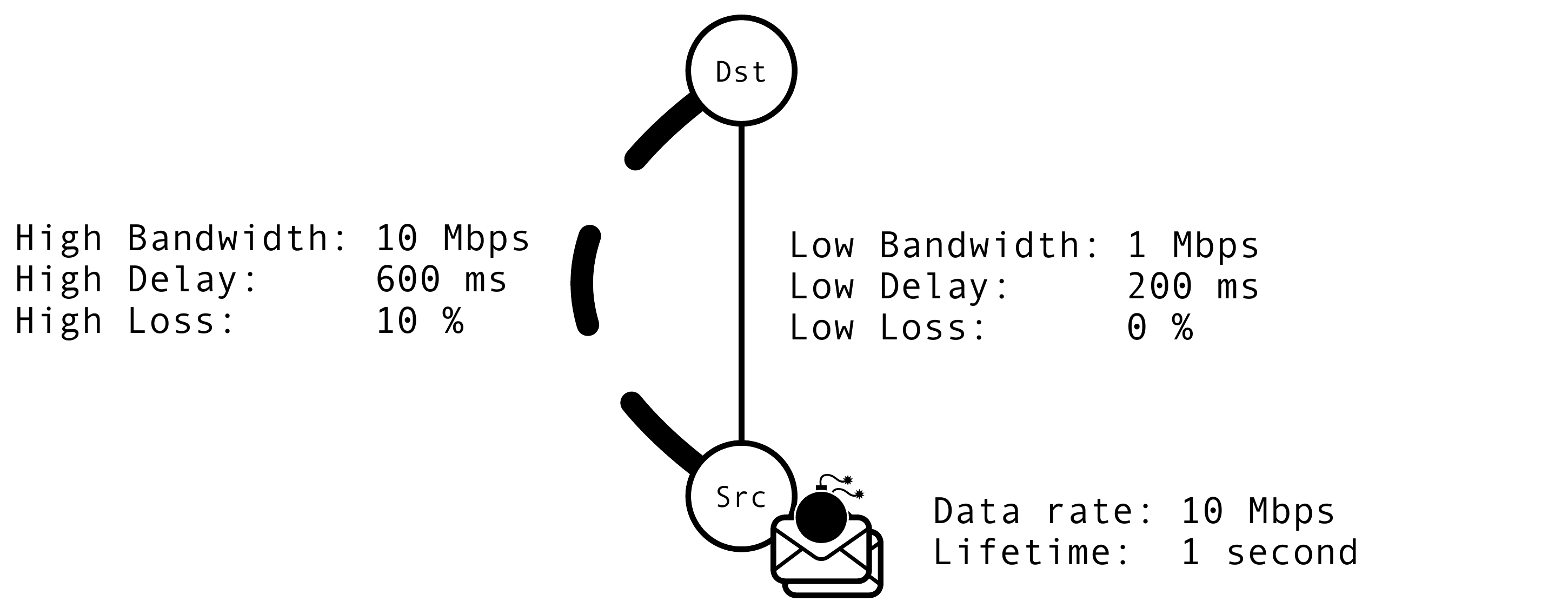}
\caption{Deadline-based multipath communication scenario.}
\label{fig:problem}
\end{figure}

Figure~\ref{fig:problem} represents a simple instance of the multipath-related
problem that we study in this paper. There are two paths with contrasting characteristics
and the source generates a constant flow of data that must be delivered, at the
latest, after one second. As the one-way delay of the high-bandwidth
path is 600~\textit{ms}, it will take 800~\textit{ms} in total for an
acknowledgment to come back along the low-latency path (assumption motivated in
Section~\ref{sec:ack}), which leaves enough time to (potentially) retransmit the data
along the low-latency path. Clearly, if all the generated data is initially sent
along the high-bandwidth path and retransmitted along the low-bandwidth path, we
can expect 100\% of the packets to reach their destination in time. This would
not be possible by using only one of the two paths.

This instance of the problem is trivial, i.e., an optimal solution can be found
intuitively. However, the problem becomes hard when more paths are considered or
when the metrics do not naturally produce such a straightforward solution. The
question we will try to answer is the following: how can the generalization of
the problem (to an arbitrary number of paths with any characteristics) be
solved?

\section{Related Work}
\label{sec:related}
Multipath TCP (MPTCP)~\cite{Raiciu2012} is the de-facto standard multipath
transport protocol. It received much attention when it was adopted by Apple for
its personal-assistant software, Siri. As MPTCP is based on TCP, it suffers from
the head-of-line blocking problem and other issues we mention in this paper;
hence it is not particularly adapted to latency-sensitive applications.
D$^2$TCP~\cite{D2Tcp2012}, on the other hand, is an example of deadline-based
protocol, but it was specially designed for data centers, not for
general-purpose settings over inter-domain networks. Moreover, D$^2$TCP is not a
multipath protocol. The partial-reliability extension of the Stream Control
Transmission Protocol (PR-SCTP)~\cite{prsctp} offers the primitive that we
examine in this paper, i.e., a possibility to define a \emph{lifetime}
parameter. Although PR-SCTP offers multihoming capabilities, additional IP
addresses are used as a backup in case of failure, so PR-SCTP is not a fully
multipath protocol and does not address the problem that we describe in this
paper~\cite{sctp}.

Diverse techniques have been used in recent work to analyze and leverage the
benefits of multipath communication. Liu et al.~\cite{Liu:2014dg} used linear
programming to evaluate multipath routing from a traffic engineering
perspective. They presented the somewhat counterintuitive result that multipath
routing offers limited gain compared to single-path routing in terms of load
balancing (under specific traffic conditions and for certain types of network
topology). However, their work---contrarily to ours---focuses on the
distribution of traffic over the network and does not take deadlines into
account. Soldati et al.~\cite{Soldati:2010eq} addressed the problem of
scheduling and routing packets with deadlines in a network whose topology is
known (represented as a directed acyclic graph), whereas we only assume
end-to-end paths. The work of Wu et al.~\cite{Chen:bn} might be the closest to
ours, but with one important difference: they propose a method to assign entire
flows (with different data rates and a deadline) to specific paths, which does
not allow using an optimal retransmission strategy. Our work falls into
another category: packet-based traffic splitting~\cite{Cetinkaya:2004kz,
Prabhavat:2011kk}. The novelty of our approach is that we leverage linear
programming to find an optimal solution to a packet-to-path assignment problem,
from the end-host's perspective, while taking cost, retransmissions, and strict
latency constraints into account. Also, we show how to integrate random delays
into our model.

\section{Background}
\label{sec:background}
In this section, we present our system assumptions and a few definitions. We
consider a network setup with a set of paths---each bearing
possibly different characteristics---between one source and one destination.
The source generates a flow of data at a constant bit rate and can split
this data and select the paths along which each part will be transmitted. In
practice, the different paths could, for example, correspond to different
network interfaces (which is the typical configuration that MPTCP
relies on~\cite{Raiciu2012}). Each bit must be delivered before a
specific point in time that we call the \emph{deadline}. To avoid any confusion,
we distinguish between a deadline, which must be interpreted as an absolute time
(e.g., 1:23:45 pm GMT), and the data's \emph{lifetime}, which must be
interpreted as a relative time (e.g., 500 ms). We consider the lifetime
to be the same for all the data, whereas the deadline depends on the lifetime
and the moment when the data was generated.

In addition to the standard bandwidth, delay, and loss characteristics of a
network path, we consider the cost of transmitting one bit along each path and set
a user-selectable upper bound on the total usage cost per unit time.
A cost can be seen, intuitively, as an amount of money that the user
must pay to utilize the path, but it can also be used to model other
consequences of using a path, such as power consumption. A system is hence
characterized by the parameters presented in Table~\ref{tab:network}.

\begin{table}[h!]
	\caption{Network Characteristics}

	\centering
	\normalsize

	\begin{tabular}{ll}
	\toprule
	& Description \\
	\midrule
	$n$ & number of independent paths \\
	$\lambda$ & data rate generated by the application \\
	$\delta$ & data lifetime \\
	$\mu$ & upper bound on total cost per second \\
	$b_i$ & bandwidth of path $i$ \\
	$d_i$ & one-way delay of path $i$ \\
	$\tau_i$ & probability of bit erasure on path $i$ \\
	$c_i$ & cost of sending one bit along path $i$ \\
	\bottomrule
	\end{tabular}

	\label{tab:network}
\end{table}

Moreover, we define $d_{\textit{min}}$ (seconds) as the shortest delay of all
paths, i.e.,
\begin{equation}
	d_{\textit{min}} = \min_{i} d_i.	
\end{equation}

Losses are modeled by a binary erasure channel. This choice is
motivated by the fact that we operate at the transport layer where checksums
are usually employed. When the verification of a checksum fails, the packet
is dropped without notifying the receiver, which is equivalent to a bit erasure.
However, we do not consider a specific packet size; instead, we use general
characteristics such as the average loss rate.

\section{Model}
\label{sec:model}
We now propose a model whose purpose is to capture the optimal multipath sending
strategy for the scenario presented above. This model can be used to provide
theoretical upper bounds on the performance of an ideal protocol under specific
conditions, but it can also be used to design an actual protocol (if combined
with different techniques and heuristics, as described in the following sections).

The problem under study is to determine what ratio of the traffic generated by
the application should be transmitted/retransmitted along each path, so that the
maximal amount of data arrives in time at the destination. Because we take
latency into account, the paths for initial transmission and for retransmission
must be considered jointly. We call this pair of transmission/retransmission
paths a \emph{path combination}. The objective is to find optimal values for the
variables contained in the following matrix:

\begin{description}
  \item $x$: matrix of size $n$-by-$n$, where $x_{i, j}$ is the proportion of
  data to send along path $i$ and then, if needed, along path $j$ (for a
  retransmission).
\end{description}

We then rearrange these variables into a vector, so that the problem can be
solved with a standard form of linear programming:

\begin{description}
  \item $x'$: vector of size $n^2$, given by the vectorization of $x$.
\end{description}

Only one retransmission is considered here in order to avoid a cumbersome
notation, but this model can clearly be adapted to an arbitrary number of
retransmissions, although the complexity of solving the problem will
naturally increase with the number of retransmissions considered, as discussed
in Section~\ref{sec:complexity}. We envision that, in most real cases, the
problem would be solved for a maximum of 2--3 retransmissions, for two reasons.
First, unless the loss rate is particularly high on all paths, having to send
the same data 4 times or more is a very rare event. Second, the time it takes to
perform many retransmissions is likely to exceed the lifetime.

\subsection{Network metrics}

We define several metrics to measure the outcomes of choosing certain values of
$x$ for a given network. This will help to define conditions and objectives in
the linear program. The metrics notation that we use is summarized in
Table~\ref{tab:metrics}.

\begin{table}[h!]
	\caption{Network Metrics}

	\centering
	\normalsize

	\begin{tabular}{ll}
	\toprule
	& Description \\
	\midrule
	$S_i$ & bit rate sent along path $i$ \\
	$G$ & goodput, i.e., useful received data rate \\
	$Q$ & communication quality (ratio of $G$ to $\lambda$) \\
	$C$ & total cost per second (sum of all paths) \\
	\bottomrule
	\end{tabular}

	\label{tab:metrics}
\end{table}

First, the amount of data sent on a certain path is obtained by considering both
the data that is sent for the first time on that path (whatever the path along
which the same data might then be retransmitted) and the data that is
retransmitted on that path (which depends on the reliability of the initial
path). Therefore, we have
\begin{equation}
\label{eq:bitrate}
  S_i = \sum_{j = 0}^{n-1} x_{i, j} \cdot \lambda
  + \sum_{j = 0}^{n-1} x_{j, i} \cdot \lambda \cdot \tau_j.
\end{equation}

This must be bounded by the available bandwidth on the corresponding path:
\begin{equation}
\label{eq:bitrate-bandwidth}
  S_i \le b_i \quad \forall~i \in \{ 0, 1, \dots, n-1 \}.
\end{equation}

Because we assume that the delay is fixed (relaxed in
Section~\ref{sec:extensions}) and that an acknowledgment always comes back on
the path with the shortest delay (discussed in Section~\ref{sec:ack}), when data
is sent along path $i$, the sender sets a \emph{retransmission timeout} to
\begin{equation}
	t_i = d_i + d_{\text{min}}.
\end{equation}

We define \emph{goodput} as the amount of application data that arrives at the
destination before the deadline each second. Again, we must consider both data
that arrives on the first attempt, and retransmitted data. As a result, the
goodput is defined as
\begin{equation}
\label{eq:throughput}
  \begin{aligned}
    G &= \sum_{i : d_i \le \delta} \sum_{j = 0}^{n-1} x_{i, j} \cdot (1 - \tau_i) \cdot \lambda \\
    &+ \sum_{i, j : d_i + d_{\textit{min}} + d_j \le \delta} x_{i, j} \cdot \tau_i \cdot (1 - \tau_j) \cdot \lambda.
  \end{aligned}
\end{equation}

Goodput depends on how much data the application generates (i.e., $\lambda$), but
we are interested in determining the proportion of $\lambda$ that a given network
can handle with an optimal strategy. Therefore, we define our main metric, which
we call \emph{communication quality}, as
\begin{equation}
\label{eq:quality}
  Q = \frac{G}{\lambda}.
\end{equation}

It follows that $0 \le Q \le 1$ (a quality of 1 meaning that all the data
arrives at the destination before the deadline).

The total cost $C$ is defined as the sum of all per-path costs per second. We
require the total cost to be bounded by a constant $\mu$. This allows us to
maximize communication quality while keeping cost to a reasonable level, but
$\mu$ can be set arbitrarily high if the cost does not have to be limited.
Therefore, the total cost is defined and bounded as follows:
\begin{equation}
\label{eq:cost}
  C = \sum_{i = 0}^{n-1} c_i \cdot S_i \le \mu.
\end{equation}

Finally, the sum of all coefficients in $x$ must equal 1. In other words, all
the bits generated by the application must be sent over the network along some
combination of paths. On one hand, we do not want to send more data than what
the application generates. On the other hand, the reason why we never send less
data is that we can achieve a more fine-grained decision of whether data should
be dropped through an additional dedicated path (the \emph{blackhole path},
presented in Section~\ref{sec:blackhole}). Therefore, we have
\begin{equation}
\label{eq:one}
  \sum_{i = 0}^{n-1} \sum_{j = 0}^{n-1} x_{i, j} = 1,
\end{equation}
\begin{equation}
  x_{i, j} \ge 0 \quad \forall~i, j \in \{0, 1, \dots, n-1\}.
\end{equation}

\subsection{Linear Programming}

We formulate our problem as a standard linear optimization:
\begin{equation}
	\label{eq:lp1}
	\begin{array}{ll}
	\text{maximize}  & p^{\text{T}} x', \\
	\text{subject to} & A x' \le q, \\
	& B x' = 1, \\
	\text{and} & x' \ge 0.
	\end{array}
\end{equation}

The objective is to maximize the communication quality $Q$, which is captured
by $p$ and follows from Equations~\ref{eq:throughput} and \ref{eq:quality}.
Each element of $p$ corresponds to a different path combination and represents
the proportion of data that can be delivered in time by using that particular
combination:
\begin{equation}
p^{\text{T}} = (p_0, p_1, \dots, p_{n^2-1}),
\label{eq:pt}
\end{equation}
\begin{equation}
p_l =
\left \{
    \begin{array}{ll}
        1 - (\tau_i \cdot \tau_j) & \text{if } d_i + d_{\textit{min}} + d_j \le \delta, \\
        1 - \tau_i & \text{if } d_i + d_{\textit{min}} + d_j > \delta \text{ and } d_i \le \delta, \\
        0 & \text{otherwise}.
    \end{array}
\right .
\end{equation}

In the above equation, $i$ and $j$ are defined as follows:
\begin{equation}
	\begin{gathered}
	  i = l \bmod n, \\
	  j = \left \lfloor \frac{l}{n} \right \rfloor.
	\end{gathered}
	\label{eq:ij}
\end{equation}

We define $i$ and $j$ from $l$ to convert indexes back to the original
(non-vectorized) notation used in $x$.

Bandwidth constraints, as presented in Equations~\ref{eq:bitrate}
and~\ref{eq:bitrate-bandwidth}, are defined by coefficients in $A$,
from $A_{0,0}$ to $A_{n-1,n^2-1}$:
\begin{equation}
A =
 \begin{pmatrix}
  A_{0,0} & A_{0,1} & \cdots & A_{0,n^2-1} \\
  A_{1,0} & A_{1,1} & \cdots & A_{1,n^2-1} \\
  \vdots & \vdots & \ddots & \vdots  \\
  A_{n-1,0} & A_{n-1,1} & \cdots & A_{n-1,n^2-1} \\
  r_0 & r_1 & \cdots & r_{n^2-1} \\
 \end{pmatrix},
\label{eq:a}
\end{equation}
\begin{equation}
A_{k, l} =
\left \{
    \begin{array}{ll}
        \lambda + \lambda \cdot \tau_i & \text{if } i = j = k, \\
        \lambda \cdot \tau_i & \text{if } i \ne k\text{, }j = k, \\
        \lambda & \text{if } j \ne k\text{, }i = k, \\
        0 & \text{otherwise.} \\
    \end{array}
\right .
\end{equation}

The cost constraint (Equation~\ref{eq:cost}) is defined with the remaining
coefficients in $A$, from $r_{0}$ to $r_{n^2-1}$:
\begin{equation}\label{eq:r_quality}
r_l = (\lambda \cdot c_i) + (\lambda \cdot \tau_i \cdot c_j),
\end{equation}
where $i$ and $j$ are defined as in Equation~\ref{eq:ij}.

The vector $q$ allows us to specify the upper bounds on bandwidth and cost:
\begin{equation}\label{eq:q_quality}
q =
 \begin{pmatrix}
  b_{0} \\
  b_{1} \\
  \vdots \\
  b_{n-1} \\
  \mu \\
 \end{pmatrix}.
\end{equation}

Finally, $B$ is used to enforce the constraint of Equation~\ref{eq:one}, i.e.,
to ensure that all the generated traffic is assigned to some path combination:
\begin{equation}
B =
 \begin{pmatrix}
  1 & 1 & \cdots & 1
 \end{pmatrix}.
\end{equation}

\subsection{Blackhole Path}
\label{sec:blackhole}

When the bitrate $\lambda$ exceeds the network's capacity, part of the data must
be dropped. The model above does not allow that situation to be represented
directly. If the sign in Equation~\ref{eq:one} was changed to $\le$, then
the total amount of data (sent on all paths combined) could be less than
$\lambda$. However, the optimal solution might consist in sending
data but not retransmitting it, which the above method does not allow.
Instead, one solution is to dedicate a \emph{virtual}
path to the function of discarding data. We refer to it as the ``blackhole
path'' and it has the following characteristics:
\begin{equation}
b_0 = \lambda, \quad
d_0 = \infty, \quad
\tau_0 = 1, \quad
c_0 = 0.
\end{equation}

Sending data along this path is equivalent to discarding the data, and this path can
be selected at any stage of (re)transmission.

\section{Extensions}
\label{sec:extensions}
Our model can be extended and transformed in several ways to specify different
objectives or consider more complex network characteristics.

\subsection{Minimizing Cost}

Instead of maximizing communication quality for a given maximum cost, it is
possible to solve the opposite problem, i.e., minimize cost for a given minimum
quality:
\begin{equation}
\begin{array}{ll}
\text{minimize}  & p^{\text{T}} x', \\
\text{subject to} & A x' \le q, \\
& B x' = 1, \\
\text{and} & x' \ge 0.
\end{array}
\end{equation}

The objective, represented by $p_l$, must be redefined as follows:
\begin{equation}
p_l = (\lambda \cdot c_i) + (\lambda \cdot \tau_i \cdot c_j).
\end{equation}

The bandwidth-related constraints (defined in $A$) and $B$ remain the same, but
the cost constraint (defined in $r_k$ and $q$) becomes a quality constraint:
\begin{equation}
r_l =
\left \{
    \begin{array}{ll}
        \tau_i \cdot \tau_j - 1 & \text{if } d_i + d_{\textit{min}} + d_j \le \delta, \\
        \tau_i - 1 & \text{if } d_i + d_{\textit{min}} + d_j > \delta \text{ and } d_i \le \delta, \\
        0 & \text{otherwise},
    \end{array}
\right .
\end{equation}
where $i$ and $j$ are defined as in Equation~\ref{eq:ij}, and
\begin{equation}
q =
 \begin{pmatrix}
  b_{0} \\
  b_{1} \\
  \vdots \\
  b_{n-1} \\
  \mu
 \end{pmatrix},
\end{equation}
where $\mu$ is, in this case, the lower bound on quality (instead of being an upper
bound on cost).

\subsection{Random delays}
\label{sec:random_delays}
Up to this point, we assumed that the delay observed on each path was constant.
Let's now consider instead that delays follow probability distributions. We denote
by $d_i$ the random variable representing the delay of a transmission on path
$i$, which follows a probability distribution $D_i$, i.e.,
\begin{equation}
\label{eq:delay-distribution}
d_i \sim D_i.
\end{equation}

One problem that arises in this situation is that the sender must determine an
additional parameter for each path combination, namely the waiting time
between the moment a packet is sent and the moment when it is possibly
retransmitted. We call this the \emph{retransmission timeout} and denote it
$t_{i, j}$.

The receiver must also choose a path (with delay $d_{\textit{min}}$) to send
acknowledgements back. We define this path to be the one with the smallest
expected delay, i.e.,
\begin{equation}
\label{eq:min}
	\textit{min} = \operatorname*{arg\,min}_{i \in \{1, \dots, n\}} \mathrm{E}[d_i].
\end{equation}

Assuming that all delays (including $d_{\textit{min}}$) are independent of each
other, the sender should find a value of $t_{i,j}$ such that
\begin{equation}
\label{eq:tij}
	t_{i, j} = \max_{t \in \mathds{R}_{+}} ( \mathrm{P}(t + d_j \le \delta) \cdot \mathrm{P}(d_i + d_{\textit{min}} \le t) ).
\end{equation}

In other words, the sender should, at the same time, find a value of $t_{i, j}$
that is small enough for the deadline to be respected, and large enough so that
a retransmission is not performed before the acknowledgement is received.

As we do not consider the duration of a transmission to be deterministic
anymore, there is now a chance that data may be correctly
transmitted/retransmitted but not respect the deadline.

We can calculate the probability of a having to retransmit a packet (sent
initially along path $i$) along path $j$ as follows:
\begin{equation}
\label{eq:proba1}
	P(\textit{retrans}_{i, j}) = 1 - P(d_i + d_{\textit{min}} \le t_{i, j}) \cdot (1 - \tau_i).
\end{equation}

Therefore, $p_l$, $A_{k, l}$, and $r_l$ become
\begin{align}
\begin{split}
\label{eq:proba2}
	p_l &= P(d_i \le \delta) \cdot (1 - \tau_i) \\
	&+ P(\textit{retrans}_{i, j}) \cdot P(t + d_j \le \delta) \cdot (1 - \tau_j);
\end{split}
\end{align}
\begin{equation}
\label{eq:proba3}
A_{k, l} =
\left \{
    \begin{array}{ll}
        \lambda + \lambda \cdot P(\textit{retrans}_{i, j}) & \text{if } i = j = k, \\
        \lambda \cdot P(\textit{retrans}_{i, j}) & \text{if } i \ne k\text{, }j = k, \\
        \lambda & \text{if } j \ne k\text{, }i = k, \\
        0 & \text{otherwise}; \\
    \end{array}
\right .
\end{equation}
\begin{equation}
\label{eq:proba4}
r_{l} = (\lambda \cdot c_i) + (\lambda \cdot P(\textit{retrans}_{i, j}) \cdot c_j).
\end{equation}

As before, $i$ and $j$ are defined according to Equation~\ref{eq:ij}.

\section{Evaluation}
\label{sec:simulation}
\subsection{Simulation Framework}
\label{sec:framework}

To demonstrate the viability of our model, we performed a series of experiments
with the \emph{ns-3} network simulator~\cite{ns3}. We set up multiple UDP
sockets between two network \emph{nodes} (a client and a server). Those sockets
are associated with different \emph{devices} (i.e., network interfaces)
communicating in pairs over a \emph{point-to-point channel}. In this setup, each
socket corresponds to a different path and we can specify the three key
characteristics (bandwidth, delay, and loss) of each one.

The client generates a total of 100,000 messages, each being 1024 bytes long
(including the application-level header), at a given constant rate. Each message
contains a simple header composed of a timestamp and a sequence number. Since
the network characteristics are all known in advance, the linear program can
directly be solved (with the CGAL library~\cite{cgal}). Thereafter, each
individual packet must be assigned to a path combination according to the
solution of the linear program. In other words, the vector $x'$ must be
discretized. The intuition behind the algorithm we present here is that we
select the path combination that will bring the actual packet distribution the
closest to the optimal solution.

We use Algorithm~\ref{algo:discretization} in the simulation framework to
determine the path combination to be used to send each packet to the server. The
algorithm works as follows. A variable maintains the total number of assignments
(i.e., the total number of packets generated so far) and an array maintains the
number of packets assigned to each combination.  At first, we select the path
combination corresponding to the highest value of $x'$, and then we select the
combination that is lacking the most packet assignments compared to the ideal
distribution of traffic represented by $x'$.

\begin{algorithm}
\SetKwFunction{select}{selectPathCombination}
\SetKwFunction{init}{initialization}
\SetKwProg{func}{Function}{}{}
\KwData{Number of paths $n$, Solution $x'$.}
\BlankLine
\texttt{// initialization} \\
\For{$i \leftarrow 0$ \KwTo $n^2-1$}{
	\texttt{assigned}[$i$] $\leftarrow 0$\;
}
$\texttt{total} \leftarrow 0$\;

\BlankLine
\func{\select{}: integer}{
	\texttt{res} $\leftarrow 0$\;
	\uIf{\texttt{total} $= 0$}{
		\texttt{res} $\displaystyle \leftarrow \operatorname*{arg\,max}_i x'_i$\;
	}
	\Else{
		\texttt{res} $\displaystyle \leftarrow \operatorname*{arg\,min}_i \frac{\texttt{assigned}[i]}{\texttt{total}} - x'_i$\;
	}
	\texttt{assigned}[\texttt{res}]++\;
	\texttt{total}++\;
	\Return{\texttt{res}}\;
}
\caption{Selection of a path combination.}
\label{algo:discretization}
\end{algorithm}

If a message is assigned to the blackhole path, then it is immediately dropped;
otherwise, the client sends it through the appropriate socket. Retransmissions
are performed after a timeout that depends on the initial path.

The server, when it receives a message, responds by sending an acknowledgment
along the lowest-delay path. The acknowledgment only contains the sequence
number of the received message. The server also verifies whether the deadline
was respected with the enclosed creation timestamp.

\subsection*{Experiment~1: Fixed Path Characteristics}

We first show simulation results in a scenario similar to the one presented in
Section~\ref{sec:problem}, i.e., with two paths that are different from each
other in every aspect. Their characteristics are defined in
Table~\ref{tab:paths}. We first assume that the network characteristics are
invariable and known by the sender, but we relax this assumption in
the next experiments.

\begin{table}[h!]
\caption{Path characteristics used in Experiments~1 and 3.}

\centering
\small
\renewcommand{\arraystretch}{1.1}

\begin{tabular}{lll}
\toprule
& Path 1 & Path 2 \\
\midrule
Bandwidth (\emph{Mbps}) $b_i$ & 80 & 20 \\
Delay (\emph{ms}) $d_i$ & 400 & 100 \\
Loss rate $\tau_i$ & 0.2 & 0 \\
\bottomrule
\end{tabular}

\label{tab:paths}
\end{table}

We simulated propagation delays of 400 and 100 milliseconds for Path~1 and
Path~2, respectively. However, queueing produces some additional delay when a
path is used at near-full capacity (see Section~\ref{sec:influence} for a
discussion on the matter). This is exacerbated by the fact that
acknowledgments, although small, are not entirely negligible. We measured in our
experiments that the deviation from the specified delay could be as high as 50
\textit{ms}. Therefore, to avoid that packets miss their deadlines by a few
milliseconds, we conservatively set delays to 450 and 150 \textit{ms} in our
model. Since the deviation from the original delay is about 50 \textit{ms} in
each direction, we set the retransmission timeouts to 100 \textit{ms} beyond the
time an acknowledgment is supposed to come back (i.e., $t_i = d_i +
d_{\text{min}} + 100 \textit{ ms}$). In practice, both propagation and queueing
delays would be reflected in RTT measurements and thus no adjustments would
be necessary.

Table~\ref{tab:solution} shows some selected solutions of the linear program for
those two paths. There are two application-related parameters that can vary: the
generated data rate~$\lambda$ and the lifetime~$\delta$. We show how the optimal
communication quality $Q$ is affected when one of those two parameters varies,
while the other is fixed.

\begin{table}[h!]
\caption{Solutions used by the sender for a network with two paths (see
Table~\ref{tab:paths}). $\delta = 800$ milliseconds (top). $\lambda = 90$ Mbps
(bottom). Columns containing only $0$'s are omitted.}

\centering
\small
\renewcommand{\arraystretch}{1.1}

\begin{tabular}{clllc}
\toprule
& \multicolumn{3}{c}{Solution} & \\
\cmidrule{2-4}
Rate $\lambda$ (\emph{Mbps}) & $x_{0,0}$ & $x_{1,2}$ & $x_{2,2}$ & Quality $Q$ \\
\midrule
10--20 & $0$ & $0$ & $1$ & $100\%$ \\
40 & $0$ & $5/8$ & $3/8$ & $100\%$ \\
60 & $0$ & $5/6$ & $1/6$ & $100\%$ \\
80 & $0$ & $15/16$ & $1/16$ & $100\%$ \\
100 & $4/25$ & $4/5$ & $1/25$ & $84\%$ \\
120 & $3/10$ & $2/3$ & $1/30$ & $70\%$ \\
140 & $2/5$ & $4/7$ & $1/35$ & $60\%$ \\
\bottomrule
\end{tabular}

\bigskip{}

\begin{tabular}{clllllc}
\toprule
& \multicolumn{5}{c}{Solution} & \\
\cmidrule{2-6}
Lifetime $\delta$ & $x_{0,0}$ & $x_{1,0}$ & $x_{1,1}$ & $x_{1,2}$ & $x_{2,2}$ & $Q$ \\
\midrule
150--400 \emph{ms} & $7/9$ & $0$ & $0$ & $0$ & $2/9$ & $22.\overline{2}\%$ \\
450--700 \emph{ms} & $0$ & $7/9$ & $0$ & $0$ & $2/9$ & $84.\overline{4}\%$ \\
750--1000 \emph{ms} & $1/15$ & $0$ & $0$ & $8/9$ & $2/45$ & $93.\overline{3}\%$ \\
1050+ \emph{ms} & $1/27$ & $0$ & $20/27$ & $0$ & $2/9$ & $93.\overline{3}\%$ \\
\bottomrule
\end{tabular}

\label{tab:solution}
\end{table}

\begin{figure}[h!]
\centering
\includegraphics[width=\columnwidth]{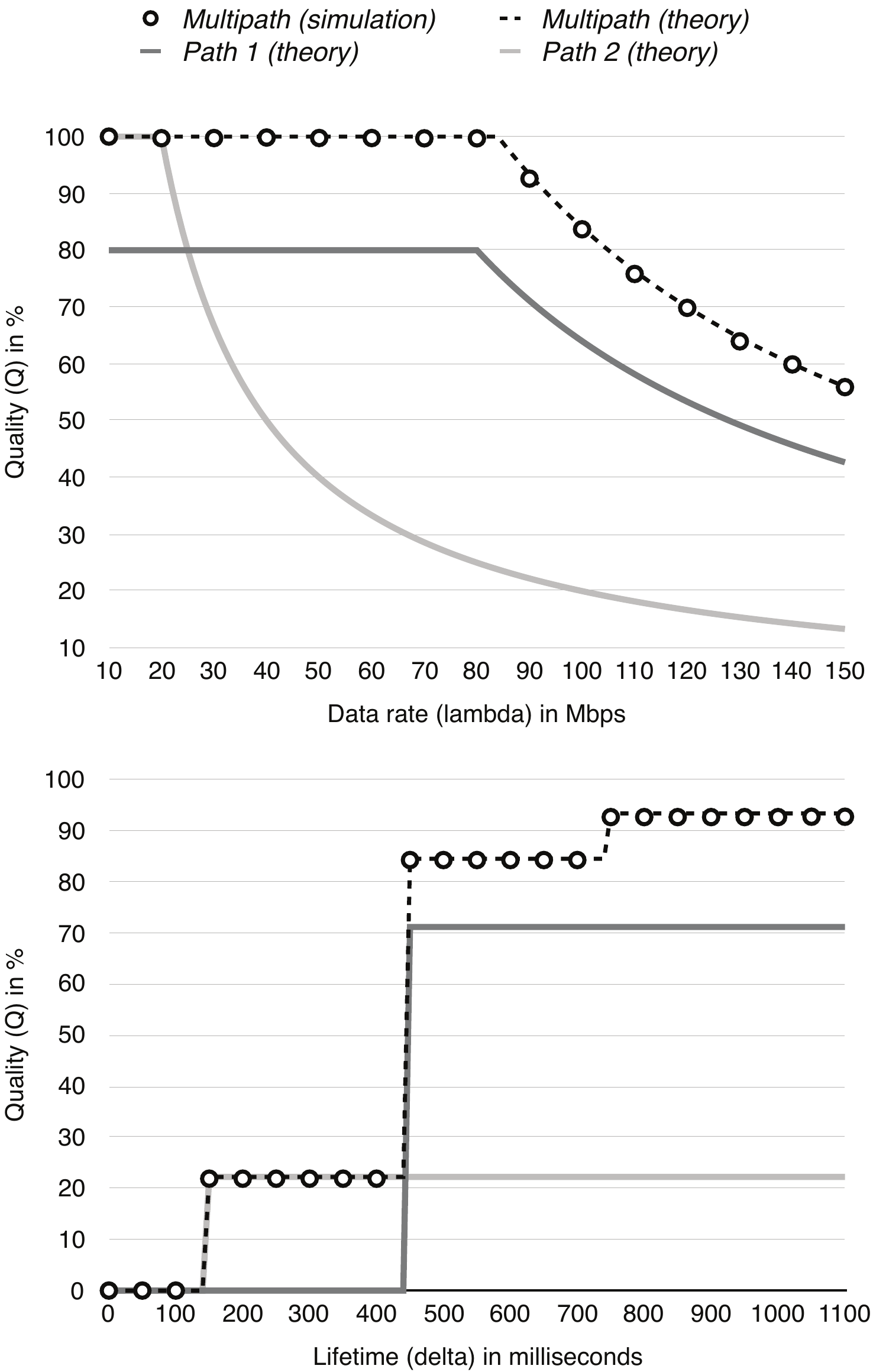}
\caption{Theoretical and simulation results for a network with two paths. See Tables \ref{tab:paths} and \ref{tab:solution} for network characteristics and solutions. $\delta = 800$ milliseconds (top). $\lambda = 90$ Mbps (bottom).}
\label{fig:res}
\end{figure}

While Table~\ref{tab:solution} shows purely theoretical performance results (and
how they can be obtained), Figure~\ref{fig:res} also shows the results of our
simulation (which closely approximates the theoretical upper bound) and the
maximum quality that can be achieved by using only one of the two paths.

\subsection*{Experiment~2: Random Delays}

In this experiment, we test the random-delay extension of our model
presented in Section~\ref{sec:extensions} in a simulation setting with two
paths. We used path characteristics similar to the first experiment, but we
added a random component to delay. It has been reported that packet delays along
a particular Internet path can be approximated by a shifted gamma
distribution~\cite{Mukherjee:1992wa, Paxson:1997td, SunggonKim:gg, Chen:2009hg}. Therefore,
we define delay on path $i$ as a random variable $d_i = X_i + \eta_i$, where
$\eta_i$ is a location parameter and where $X_i$ is a random variable following
a gamma distribution, i.e., with the following cumulative distribution function:
\begin{equation}
	P(X_i \le x) = \frac{\gamma(\alpha_i, \beta_i x)}{\Gamma(\alpha_i)},
\end{equation}
where
\begin{equation}
	\Gamma(\alpha) = \int_{0}^{\infty} x^{\alpha-1}e^{-x} dx
\end{equation}
and
\begin{equation}
	\gamma(\alpha, x) = \int_{0}^{x} t^{\alpha-1} e^{-t} dt.
\end{equation}

Such a random variable has an expected value
$\mathbb{E}[d_i] = \eta_i + \alpha_i \beta_i$ and a variance
$\text{Var}[d_i] = \alpha_i \beta_i^2$.

The distribution parameters and the other network characteristics used in this
simulation are shown in Table~\ref{tab:random_delays}. To minimize the effects
of queueing delay and concentrate on the simulated delay distribution, we
over-provisioned both paths (in terms of bandwidth), but only used the allowed
amount of bandwidth ($b_i$) specified in the model (see
Section~\ref{sec:influence} for a discussion on queueing delays).

\begin{table}[h!]
\caption{Path characteristics used in Experiment~2.}
	\centering
	\small
	\renewcommand{\arraystretch}{1.1}
	\begin{tabular}{lll}
		\toprule
		& Path 1 & Path 2 \\
		\midrule
		Bandwidth (\emph{Mbps}) $b_i$ & 80 & 20 \\
		Delay (\emph{ms}) parameter $\eta_i$ & 400 & 100 \\
		Delay (\emph{ms}) parameter $\alpha_i$ & 10 & 5 \\
		Delay (\emph{ms}) parameter $\beta_i$ & 4 & 2 \\
		Loss rate $\tau_i$ & 0.2 & 0 \\
		\bottomrule
	\end{tabular}
\label{tab:random_delays}
\end{table}

To calculate retransmission timeouts, we use Equation~\ref{eq:tij} that we can
rewrite here as
\begin{align}
\begin{split}
	t_{i, j} = \max_{t \in \mathds{R}_{+}} ( &F_{X_j}(\delta - \eta_j - t) \\
	&\cdot (F_{X_i}(t - \eta_i) * f_{X_{\text{min}}} (t - \eta_\text{min})) ),
\end{split}
\end{align}
where $F_X()$ and $f_X()$ denote, respectively, the \emph{cumulative
distribution function} and the \emph{probability density function} of a random
variable $X$, and $*$ stands for convolution. The above method does not
necessarily produce a unique solution. In this case, the optimal timeouts that
we choose are
\begin{equation}
	\begin{gathered}
		t_{1, 2} = 615 \textit{ ms}, \\
		t_{2, 1} = 252 \textit{ ms}, \\
		t_{2, 2} = 323 \textit{ ms}.
	\end{gathered}
\end{equation}

The timeout $t_{1, 1}$ is not defined here because it is not possible to perform
a retransmission in time with that particular path combination and a lifetime of
750 \textit{ms}.

In this network setting, when we generate data at a rate $\lambda =
90$ Mbps, with a lifetime $\delta = 750$ ms, our model extension indicates that
the expected quality is $93.\overline{3}\%$ and when we used the extension in
the simulation, out of 100,000 generated packets, 93,332 were received before
their deadline. This indicates that the model produces realistic results and
that Algorithm~\ref{algo:discretization} closely approximates theoretical
values in the long run.

\subsection*{Experiment~3: Sensitivity}

\begin{figure}[h!]
\centering
\includegraphics[width=\columnwidth]{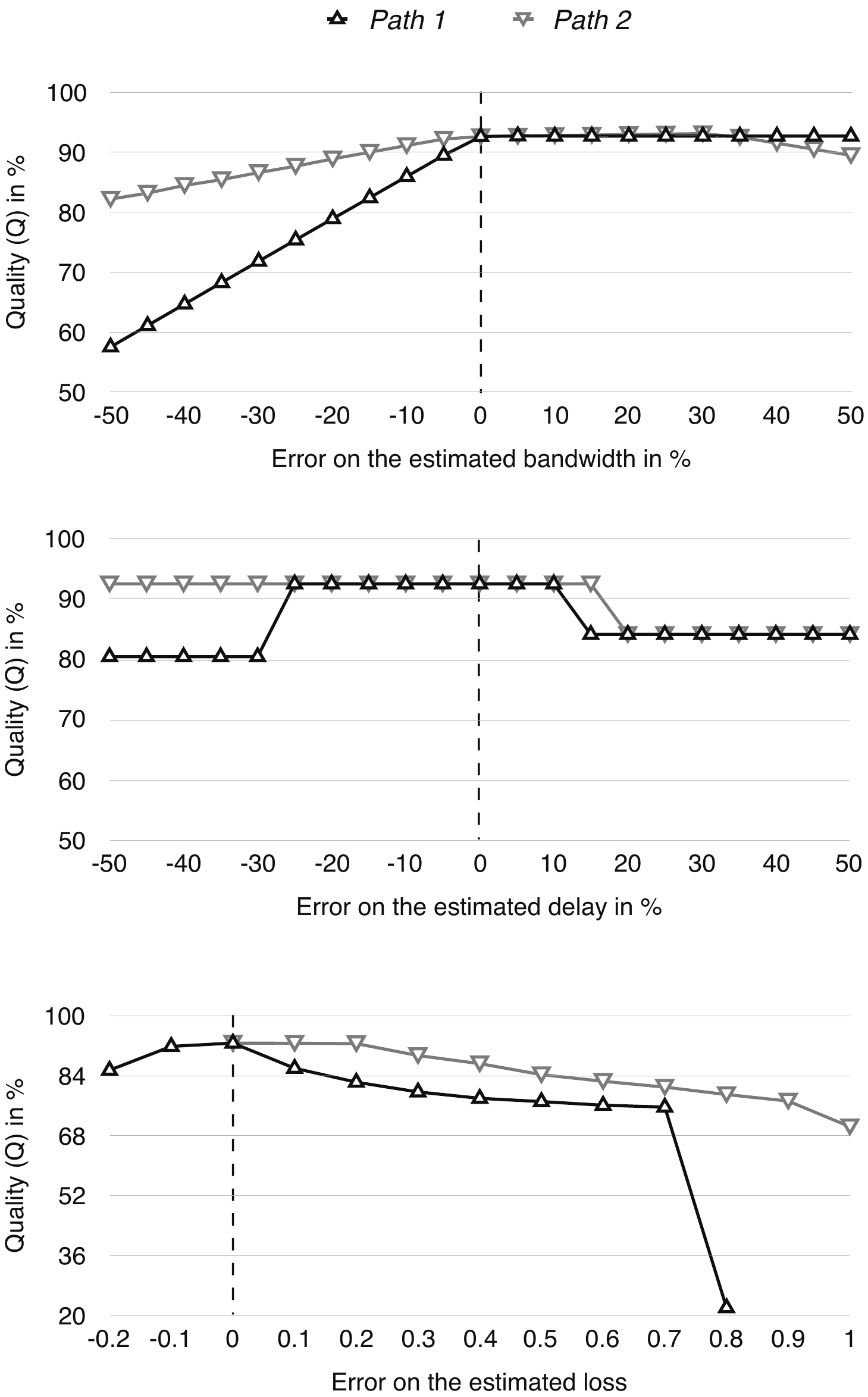}
\caption{Simulation results showing the performance of our multipath model in function of estimation errors on the two different paths.}
\label{fig:sensitivity}
\end{figure}

Since our model requires the sender to estimate the end-to-end characteristics
of the network being used (discussed in Section~\ref{sec:practice}), we analyse
how sensitive the model is to inaccurate estimations.
In particular, Figure~\ref{fig:sensitivity} (top) shows how erroneous bandwidth
estimation affects the communication quality when two paths are used
simultaneously (with the network characteristics presented in
Table~\ref{tab:paths}, $\lambda = 90$ Mbps, $\delta = 800$ milliseconds).
On the left-hand side of the vertical dashed line, if the capacity
of the network is underestimated,
unsurprisingly, the quality decreases because the model forces
packets to be dropped. On the right-hand side, however, messages are not
dropped and congest the network. Therefore, the loss rate increases
proportionally due to overflowing packet buffers and the communication
quality remains mostly unaltered.

With regard to delay (middle of Figure~\ref{fig:sensitivity}), as expected, the
quality is maximal when there is no estimation error. Moreover, there is a large
plateau at the maximum quality value, which indicates that, in this particular
scenario, the model is not sensitive to small ($<10$\%) erroneous delay
estimations.

Finally, the bottom part of Figure~\ref{fig:sensitivity} shows that erroneously
estimating loss (by a reasonable amount) also results in a small decrease in
communication quality.

\section{Practical Considerations}
\label{sec:practice}
\subsection{Estimation Techniques}

The model presented above assumes that the sender has a some knowledge about the
network's characteristics. To use the model in practice, it is necessary to
estimate bandwidth, delay, and loss, on each path. In this section, we discuss
approaches for estimating these values in a real-world setting.

\paragraph{Bandwidth Estimation}
The bandwidth of a network path is probably the most challenging metric to
estimate, for several reasons. Firstly, bandwidth is a broad term and can refer
to at least three different specific metrics in the context of data networks:
\emph{capacity} (maximum possible bandwidth), \emph{available bandwidth}
(maximum unused bandwidth), and \emph{TCP throughput} or \emph{bulk transfer
capacity} (throughput obtainable by a single TCP connection, which is not an
appropriate metric in the context of this paper)~\cite{Prosad2003}. Moreover,
for each of these metrics, several estimation techniques have been proposed and
many tools (open~source or commercial) are available, each with advantages and
drawbacks.
Another important aspect to consider is congestion control. TCP typically uses
window-based congestion control, but other schemes based on an explicit
optimization of the sending rate have been developed. For example,
PCC~\cite{Dong:2014} adjusts the sending rate depending on the outcome of a
utility function. When the system reaches a stable state, the rate determined by
the congestion control algorithm can directly be used as the value of $b_i$ in
our model.

\paragraph{Delay Estimation}
Estimating the average delay is relatively straightforward. As soon as an
acknowledgment is received, an RTT value can be computed. However, as we assume
that acknowledgments always come from the same path (the one with the lowest
latency), estimating the delay of all paths requires a more elaborate
acknowledgment scheme (such as the one we outline in Section~\ref{sec:ack}).
To estimate the probability distribution $D_i$ that delay follows on 
a given path (Equation~\ref{eq:delay-distribution}), two approaches are
possible. First, if a specific distribution is assumed (e.g., a shifted gamma
distribution), then its parameters
can be estimated through regressions analysis~\cite{Chen:2009hg}.
Alternatively, the problem can be discretized by recording a sample of packet
delays to determine average delays in place of expected values
(Equation~\ref{eq:min})
and discrete probability distributions instead of continuous ones
(Equations~\ref{eq:tij}--\ref{eq:proba4}).

\paragraph{Loss Estimation}
The loss rate of a path is estimated by dividing the number of lost packets by
the total number of packets sent along that path. To obtain an accurate
estimation, this process requires that a large number of packets have been sent.
For that reason, the loss rate can first be set to 0\% and the sending strategy
can then be refined every time a loss is recorded.

\subsection{Complexity}
\label{sec:complexity}

There exists a profusion of libraries in various languages to solve linear
programs. The time complexity of the employed algorithm is not of critical
importance for the usage of our model in a protocol (since the problem size is
relatively small in most practical scenarios), as long as the metric estimations
are stable.

We stress that it is not necessary to solve the problem for every packet, but
only when the estimations of network characteristics vary significantly. Indeed,
in the best case, the problem must be solved only once---as soon as all metrics
are available, and not again if metrics remain stable. However, if metrics
experience volatility, or if paths go up/down, then it is important to show that
solving our linear program does not constitute a heavy computational burden.

Real-valued linear programs can be solved in (worst-case) polynomial time in
terms of the number of variables, with different variants of the interior-point
method. However, in the case of Equation~\ref{eq:lp1}, the size of $x'$ grows
exponentially with the number of retransmissions considered. To be precise, for
a problem with $n^m$ variables ($n$ being the number of paths and $m$ the number
of retransmissions) and that can be encoded in L input bits, Karmarkar's
algorithm~\cite{Strang:1987jt} requires $O(n^{(7/2)m}L)$ operations.

\begin{figure}[h!]
\centering
\includegraphics[width=\columnwidth]{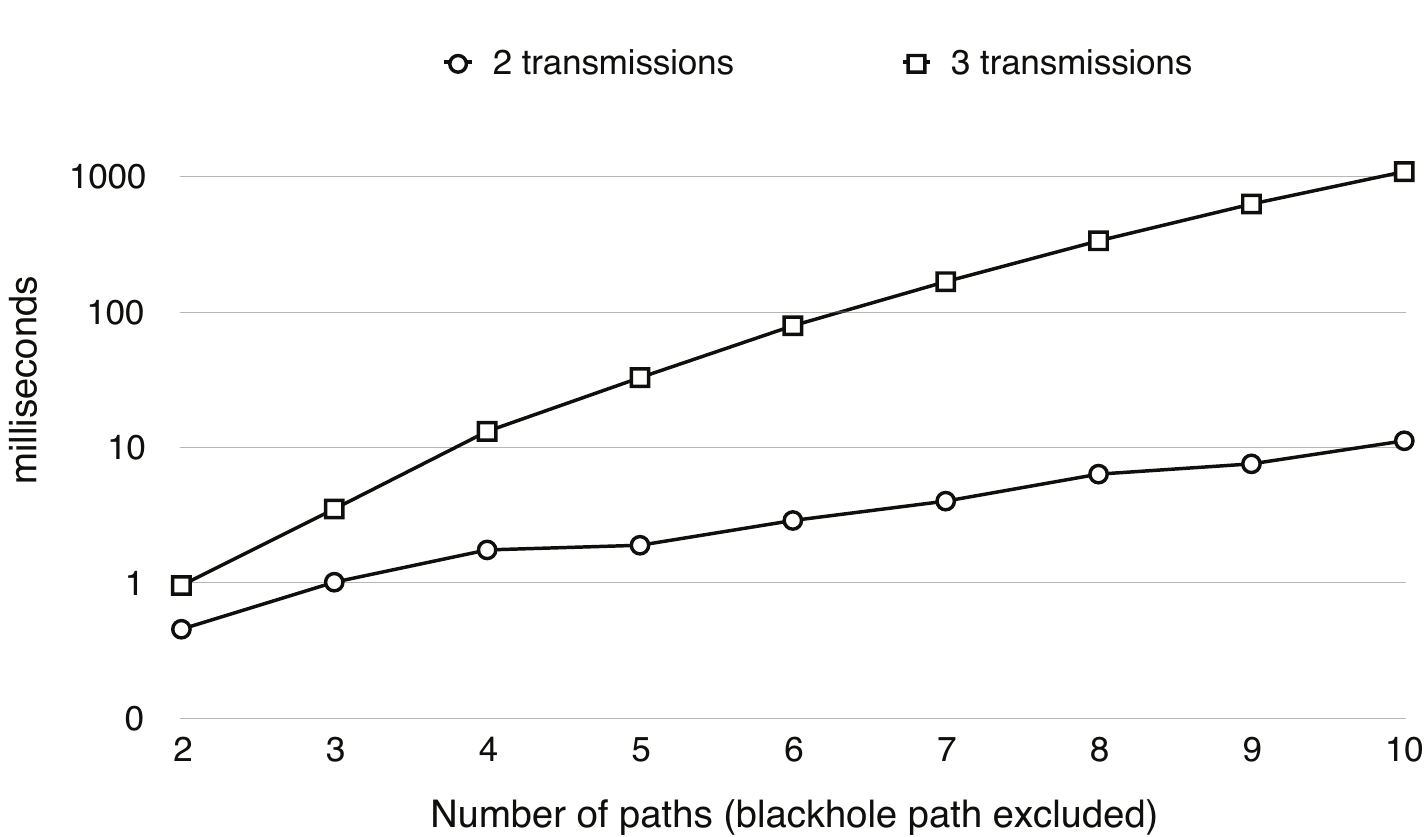}
\caption{Computation times for solving multipath problems of different sizes (linear programming). The y-axis is in logarithmic scale.}
\label{fig:complexity}
\end{figure}

Our experiments with a commodity machine (2.8 GHz Intel Core i5, 8 GB 1600 MHz
DDR3) and the CGAL library~\cite{cgal} show that, on average (calculated over
100 runs), it takes about 458.39 microseconds to solve a problem in which we
consider two paths (excluding the blackhole path) and two transmissions per data
unit, for example, which is negligible given that solving the problem does not
block packet transmissions. Figure~\ref{fig:complexity} shows computation times
for larger problems (averaged over 100 runs).

\subsection{Acknowledgment scheme}
\label{sec:ack}
In our model, we assume that acknowledgments cannot get lost, always take the
lowest-latency path, and yet that RTT estimation is possible. One important
observation is that all the data should be acknowledged through the same path
the data came from for precise and accurate RTT estimation. However, this does
not mean that the acknowledgment cannot also contain information about other
packets that were (or were not) received on other paths.

To get as close as possible to the above-mentioned assumption in practice, the
acknowledgments sent in response to every (or every $n$) packet(s) should
contain a combination of the following pieces of information:
(a) the range of (i.e., the lowest and the highest) packet numbers that the receiver is expecting,
(b) a bit vector and its position indicating what was already received in a set of consecutive packets,
(c) the packet that was just received (for RTT estimation).

In links with low bandwidth-delay products, an acknowledgement packet may
contain enough information to describe the entire set of packets that are
in-flight between the sender and receiver. However, when the bandwidth-delay
product is large, and the lowest-latency path is lossy, the design of such
acknowledgements is more complicated. Specifically, the bit vector indicating
which packets have been received and which packets have not been received may be
shorter than the packets in flight, for reasons such as maximum packet size and
the desire to reduce overhead. In such cases, the receiver's acknowledegment
scheme becomes an integral part of reaching the desired quality metric. The goal
is to create an acknowledgement stream that maximizes the quality for a given
cost. However, because we do not know which acknowledgement packets will be
lost, we can only maximize the expected quality, and because our acknowledgement
algorithm must be nearly on-line (and therefore does not have knowledge of
future acknowledgement transmission times), such quality can only be optimized
with respect to a particular timing for future acknowledgements. We leave to
future work the problem of designing a high-performance, low-overhead
acknowledgement scheme that performs well for both low and high bandwidth-delay
products.

\subsection{Retransmissions}

In addition to using a retransmission timeout as described in
Section~\ref{sec:random_delays}, it is possible to implement a
\emph{fast-retransmission} mechanism (similar to TCP's ``fast retransmit''
enhancement~\cite{rfc2581}), based on the fact that per-path packet re-ordering
is a relatively unlikely event in the communication architecture we consider.
This allows correcting for inappropriate timeout values caused by erroneous
delay estimations, when the amount of generated traffic is sufficient.

In TCP, the mechanism is triggered after three duplicate acknowledgments, but no
formal motivation is provided for this particular number. Therefore, the
question of exactly how such a mechanism should work in our context remains
open.

\section{Discussion}
\label{sec:discussion}
\subsection{Path Characteristics Influenced by Usage}
\label{sec:influence}
In some cases where a path has limited resources (such as bandwidth and queue
length) relative to our ability to use those resources, our usage of a path may
impact the performance characteristics of that path. A mostly-saturated link,
when it encounters increased traffic, may exhibit a higher loss rate than the
one initially measured; likewise, queuing theory shows that as utilization
increases,  latency also increases. These effects introduce non-linearities in
our model, since changes in $x$ affect latency and loss rates, and thus quality
$p^{\text{T}}$ (Equation~\ref{eq:pt}) and bandwidth usage $A$
(Equation~\ref{eq:a}).
In such environments, we can initially assume that the characteristics of each
path are independent of transmission rate. As long as our path usage does not
change, there is no impact on our linear-programming solution. Otherwise, we
gather link characteristic information as the path usage changes and determine
whether a statistically significant change occurs  in link characteristics. If
so, we model the link's latency and loss  as a function of input bandwidth, and
replace Equation~\ref{eq:lp1} with a non-linear program  that takes into account
the impact of transmission rates on quality and bandwidth limits.

When two paths share a common subpath, traffic sent on one path can influence
the properties of traffic sent along the other path. In many network
architectures,  we can determine that two paths are linked in this way; for
example, on the Internet, traceroute may reveal  the extent to which two paths
share a common subpath. Detecting such situations and modifying the non-linear
program appropriately is beyond the scope of this paper, and left as future
work.
Conveniently, our algorithm tends to send packets smoothly; that is, the
inter-arrival time between  two consecutive packets sent on the same path tends
to be similar as long as the sending rate is smooth. As a result, our approach
need not consider the impact of traffic patterns (rather than traffic amounts)
on queuing latency and loss.

\subsection{Channel Coding}

Our model does not include any form of channel coding and focuses instead on the
optimal transmission/retransmission strategy. It is established that correlated
losses decrease the effectiveness of open-loop error control schemes (such as
forward error correction) and experiments showed that losses are correlated even
when as little as 10\% of capacity is used~\cite{Bolot93}. Although there might
be an opportunity to de-correlate losses by sending consecutive packets along
different paths, that approach has limitations. When a packet is lost, the delay
required to recover the corresponding group of packets equals the longest delay
of all paths. Therefore, the benefits of end-to-end coding (including in a
multipath context) are questionable and need to be further investigated.
Moreover, in terms of fairness to other users/applications, only performing
retransmissions with no additional redundancy (due to coding) is more desirable.

\subsection{Interpretation of Bandwidth and Cost Limits}

Our model operates on the expected value of the bandwidth to be used and the
cost bound.  In particular, the values in $A$ (Equation~\ref{eq:a}) use the
traffic vector to calculate  the expected usage of each link and the expected
cost. In some systems, exceeding a user-specified cost bound may be
unacceptable; in other systems, exceeding the pre-specified bandwidth limits
may result in packet loss that cannot be handled by our model. In such cases, a
system using our approach  can adjust the values in $q$
(Equation~\ref{eq:q_quality}) until an acceptable solution is reached. In
particular, given a certain data rate, number of packets, and rate solution
$x'$, we can compute the probability of exceeding an expected cost  or a
bandwidth limit; in the event that this probability is too high, the system can
adjust the bandwidth limit or cost limit and re-solve the linear program to
obtain a solution that is closer to the system's goals.

\section{Conclusion}
\label{sec:conclusion}
Packet switching has the advantage of enabling the usage of several network
paths simultaneously for a single stream of data or even for a single message,
which constitutes an attractive research area for improving network
performance. Unfortunately, the path diversity that the Internet offers is
rarely fully exploited, for several reasons. First, the current Internet
architecture does not allow end hosts to specify the path(s) they want to use.
Second, the multipath paradigm poses many new challenges and requires that most
transport-layer concepts be redesigned. Finally, the advantages that multipath
communication offers are not well known because they have not yet been
sufficiently examined.

In this paper, we proposed an analytical model for optimizing the performance of
partially-reliable multipath communications, in particular with the goal of
developing better protocols for latency-sensitive applications. We showed that
path diversity achieves better performance than uniform paths in deadline-based
scenarios, through theoretical and simulation results. Many challenges remain to
design a deployable protocol (e.g., cross traffic, varying conditions,
congestion/flow control), which we leave to be addressed by future work.
However, multipath communication promises to provide a multitude of desirable
properties.

\section*{Acknowledgments}
\label{sec:acknowledgments}
The research leading to these results has received funding from the European
Research Council under the European Union's Seventh Framework Programme
(FP7/2007-2013), ERC grant agreement 617605.
This material is also based upon work partially supported by NSF under Contract No.
CNS-0953600. The views and conclusions contained here are those of the authors
and should not be interpreted as necessarily representing the official policies
or endorsements, either express or implied, of NSF, the University of Illinois,
or the U.S. Government or any of its agencies.

\balance

\bibliographystyle{IEEEtran}
\bibliography{ref}

\end{document}